\documentclass[%
reprint,
superscriptaddress,
amsmath,amssymb,
aps,
pra,
]{revtex4-2}

\usepackage{graphicx}
\usepackage{dcolumn}
\usepackage{bm}
\usepackage{hyperref}
\usepackage{microtype}
\usepackage{ragged2e}
\usepackage{physics}
\usepackage{nicefrac}
\newcommand{\Ai}{\operatorname{Ai}}

\newcommand{\dif}{\mathrm{d}}

\hypersetup{
     colorlinks=true,
     linkcolor = blue,
     citecolor = blue,
     urlcolor = black
}

\begin{document}

\title{Dynamics of self-accelerating electron beams in a homogeneous magnetic field}

\author{Michael Goutsoulas}

\email{goutsal@math.uoc.gr}

\affiliation{Department of Mathematics and Applied Mathematics, University of Crete, 70013 Heraklion, Crete, Greece}

\author{Nikolaos K. Efremidis}



\affiliation{Department of Mathematics and Applied Mathematics, University of Crete, 70013 Heraklion, Crete, Greece}

\affiliation{Institute of Applied and Computational Mathematics, Foundation for Research and Technology-Hellas (FORTH), 70013 Heraklion, Crete, Greece}

\date{\today}

\begin{abstract}
  We examine the dynamics of electron beams that, in free space, are self-accelerating, in the presence of an additional magnetic field. We focus our attention in the case of Airy beams that follow parabolic trajectories and in generalized classes of beams associated with power-law trajectories. We study the interplay between beam self-acceleration and the circular motion caused by the magnetic field. In the case of Airy beams, using an integral representation, we find closed-form solutions for the electron wavefunction. We also derive asymptotic formulas for the beam trajectories both for Airy beams and for self-accelerating power-law beams. A ray optics description is rather useful for the interpretation of the beam dynamics. Our results are in excellent comparison with direct numerical simulations. 
\end{abstract}

\maketitle

\section{INTRODUCTION}

In quantum mechanics, Airy wave packets were proposed as solutions of the potential-free Schr\"odigner equation~\cite{berry-ajp1979}. Such probability waves are associated with some very unique features. In particular, Airy waves are the only diffraction-free solutions of the Sch\"odinger equation with one spatial dimension~\cite{unnik-ajp1996}, are self-healing~\cite{broky-oe2008}, and propagate along a self-accelerating parabolic trajectory~\cite{berry-ajp1979}.

Although Airy wave packets initiated from the area of quantum mechanics, such self-acceleraging waves found a fertile ground in optics, mainly due to the plethora of associated applications~\cite{efrem-optica2019,hu-springer2012}.
Self-accelerating Airy beams in optics were proposed and observed in~\cite{sivil-ol2007,sivil-prl2007}. In these works it was shown that an exponentially truncated Airy beam, which has finite power and thus is experimentally realizable, can be easily generated in the Fourier space by applying a cubic phase mask to a Gaussian beam. Subsequently, self-accelerating beams following different trajectories have been examined both in the paraxial~\cite{green-prl2011,chrem-ol2011} and in the non-paraxial domain~\cite{froeh-oe2011,kamin-prl2012,gouts-pra2018}. The properties of Airy, accelerating, and abruptly autofocusing beams~\cite{efrem-ol2010}, are utilized in different applications in optics, ranging from particle manipulation~\cite{baumg-np2008,zhang-ol2011}, ablation~\cite{papaz-ol2011} and micromachining~\cite{mathi-apl2012}, imaging~\cite{jia-np2014,vette-nm2014}, filamentation~\cite{polyn-science2009} and electric discharge generation~\cite{cleri-sa2015}, and plasmonics~\cite{salan-ol2010,minov-prl2011,zhang-ol2011-plasmon,li-prl2011}. The dynamics of Airy beams in the presence of an additional dynamic linear potential~\cite{efrem-ol2011} and a parabolic potential~\cite{zhang-oe2015,zhang-ol2015} have also been studied.

Thirty four years after their theoretical prediction~\cite{berry-ajp1979}, self-accelerating quantum free electron wave packets of the Airy type were experimentally observed in~\cite{voloc-nature2013}, by diffraction of electrons through a nanoscale hologram. Self-accelerating Dirac particles were studied in~\cite{kamin-np2015}. It was shown that engineering the wave function of electrons, as accelerating shape-invariant solutions to the potential-free Dirac equation, fundamentally acts as a force. 

For electron beams, we know that the presence of a constant magnetic field gives rise to a rotational motion of the position expectation value with constant radius and angular frequency~\cite{thaller-2000visual}. The question that we would like to examine is how such a circular accelerating motion is affected by the presence of a wave packet that is self-accelerating.

In this work, we examine the dynamics of electron wave packets under the combined action of self-acceleration due to the initial condition, and the circular accelerating motion due to a constant magnetic field. In the case of beams of the Airy type, we utilize the Mehler kernel to derive the dynamics in closed-form. From the exact solution, we can also derive the trajectory of the electron wave. In addition, we study the dynamics of wave packets, which in the absence of a magnetic field, follow a power-law self-accelerating trajectory. The collective behavior of such beams due to self-acceleration and the constant magnetic field on the beam trajectory, is determined by utilizing integral asymptotics. The origin of the resulting motion is unveiled by utilizing a ray optics description. We perform a series of numerical simulation which show that our theoretical predictions are in very good agreement with the numerical results.

\section{Paraxial dynamics of electron beams in a constant magnetic field}
The dynamics of an electron wave packet in the presence of an electromagnetic field is given by
\[
  \left(i\hbar\frac\partial{\partial t}
    +e\phi(\bm r,t)
  \right)\Psi
  =
  \frac1{2m}\left(\hat{\bm p}+\frac ec\bm A
  \right)^2\Psi,
\]
where $\Psi$ is the wavefunction, $\bm A$ and $\phi$ are the vector and the scalar potentials, $\hbar$ is Planck's constant, $m$ is the mass, $e$ is the electron charge, and $\hat{\bm p}=-i\hbar\nabla$ is the momentum operator. We select a constant magnetic field along the propagation $z$-direction $\bm B = \nabla\times\bm A = B\bm e_z$, and a symmetric gauge $\bm A = -(1/2)\bm r\times\bm B=-(B/2)(y\bm e_x-x\bm e_y)$ with $\phi=0$. We express the wave packet as 
\[
  \Psi = \psi e^{i(p_0z-E_0t)/\hbar},
\]
where $E_0=p_0^2/(2m)$, and define a coordinate system that moves along with the wave packet $\zeta=z-\hbar k_0t/m$, $\tau=t$. We use normalized dimensionless coordinates $z\rightarrow z_0z$, $t\rightarrow t_0t$, $(x,y)\rightarrow L(x,y)$, and $B\rightarrow B_0B$ with $t_0=mL^2/\hbar$, $z_0=k_0L^2$, $B_0=c\hbar/(eL^2)$, and $p_0=\hbar k_0$. In the paraxial regime we can ignore $\psi_{zz}$ and thus obtain
\begin{equation}
  i\partial_\tau\psi= 
  \left[
  -\frac12\nabla_\perp^2 
  -i\frac B2(x\partial_y-y\partial_x)
  + 
  \frac{B^2}8(x^2+y^2)
  \right]\psi,
  \label{eq:schrodinger}
\end{equation}
where $\nabla_\perp=\partial_{xx}+\partial_{yy}$ is the Laplacian in the transverse plane. The integral representation of Eq.~(\ref{eq:schrodinger}) is provided by
\begin{equation}
  \psi(\bm r,\tau) = \iint\psi_0(\bm\rho)K(\bm r,\bm\rho,\tau)\,\dif\xi\dif\eta,
\end{equation}
where
\begin{equation}
  K=
  \frac{B}{4\pi i \sin\frac {B\tau}2}
  e^{i\frac B4\cot\frac{B\tau}2(\bm r-\bm\rho)^2+
    i\frac B2\bm e_3\cdot(\bm r\times\bm\rho)}
  \label{eq:K}
\end{equation}
is the respective Mehler kernel, $\bm r = x\hat{\bm e}_x+y\hat{\bm e}_y$, and $\bm\rho=\xi\hat{\bm e}_x+\eta\hat{\bm e}_y$. Note that Eq.~(\ref{eq:K}) can be derived from the Mehler kernel of the two-dimensional harmonic oscillator problem, by applying a rotation that accounts for the action of the orbital angular momentum operator~\cite{thaller-2000visual,galla-pra2012}. Expanding the vector identities we obtain the following more convenient form for our calculations 
\begin{equation}
  \psi= \frac{w}{2\pi i} \iint_{\mathbb R^2}\psi_0(\xi,\eta)
  e^{iv (\xi^2+\eta^2)-i(S_1 \xi + S_2 \eta)}\dif\xi \dif\eta,
\label{Sol2}
\end{equation}
where in the above expression
$v=(B/4)\cot(B\tau/2)$,
$w= (B/2)\csc(B\tau/2)e^{iv(x^2+y^2)}$,
$S_1=2v(x+y\tan(B\tau/2))$,
and
$S_2=2v(y-x\tan(B\tau/2))$. From Eq.~(\ref{Sol2}) we see that the electron wave $\psi$ can be considered as the two-dimensional Fourier transform of the initial condition multiplied with a parabolic phase term $\psi_0(\xi,\eta)e^{iv(\xi^2+\eta^2)}$, with $S_1$ and $S_2$ being the spatial frequencies conjugate to the $\xi$ and $\eta$ variables, respectively.

\begin{figure}
  \centering
  \includegraphics[width=0.6\columnwidth]{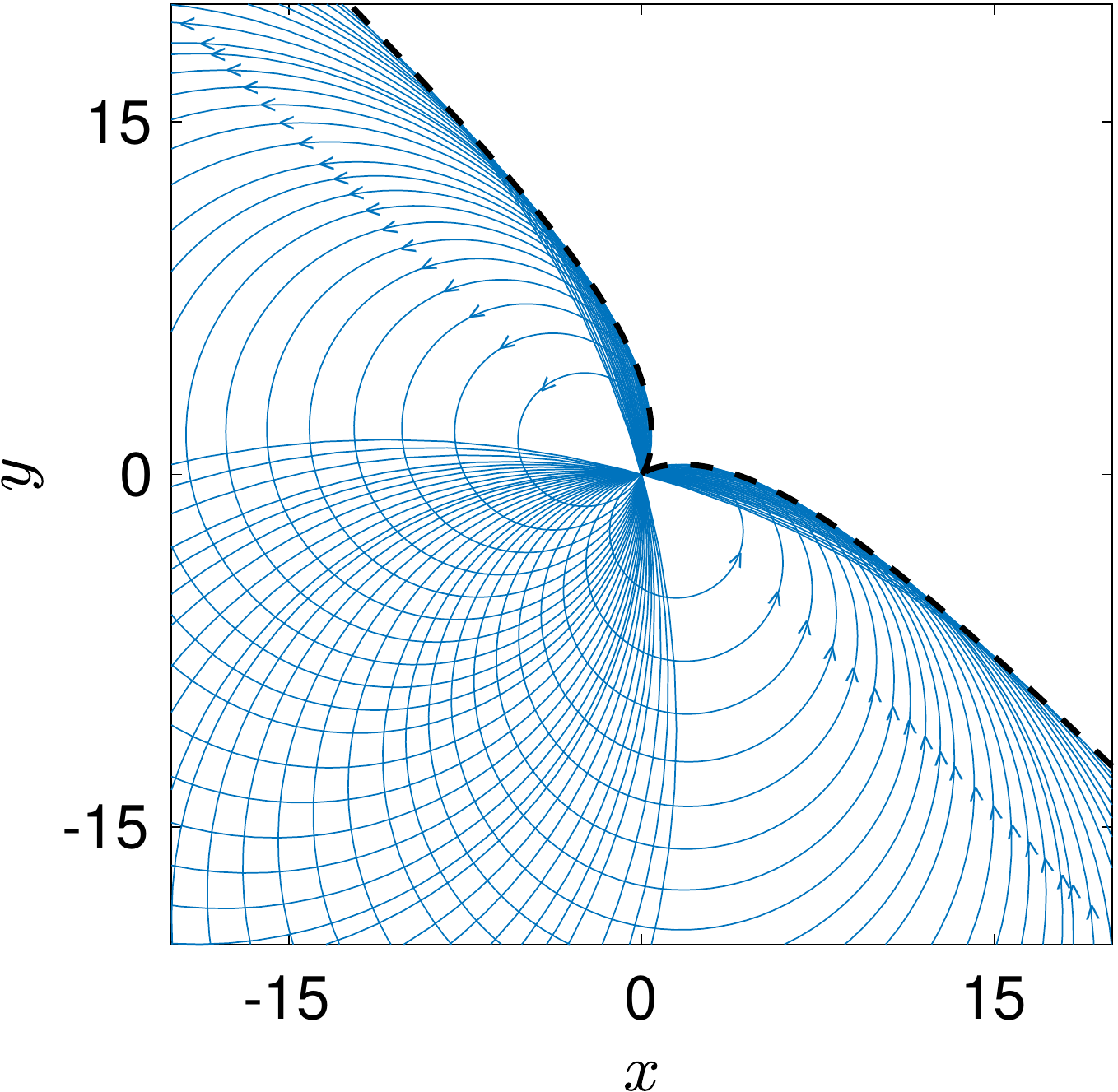}
  \caption{Ray picture of an electron Airy beam propagating in a constant magnetic field on the transverse $x-y$ plane. The propagation direction, $z$, is the same as the direction of the magnetic field. The rays are shown as blue circles while the resulting cusp caustic is shown with a black dashed curve.\label{fig:ray_picture_Airy}}
\end{figure}

\begin{figure*}
  \centering
  \includegraphics[width=\textwidth]{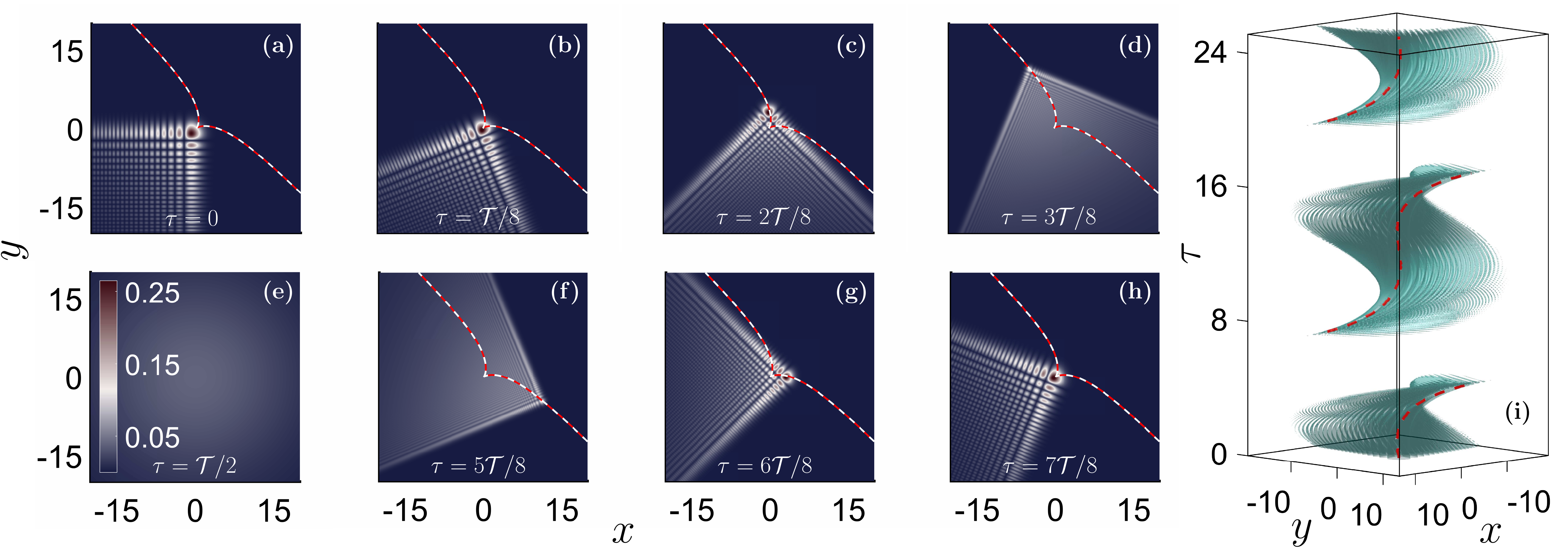} 
  \caption{Dynamics of an electron Airy beam in a constant magnetic field with $B=1/2$. The electron wave at the input plane is given by Eq.~(\ref{eq:Airy}) with $\kappa_1=\kappa_2=1$, $\alpha=0.04$, and $\beta_1=\beta_2=0$. Cross sections of the wave amplitude $|\psi|$ are shown on the left panel in multiples of $1/8$th of the period. The red-white dashed line is the theoretical prediction for the Airy beam trajectory. The colormap shown in (e) is the same for all the cross-sections. In (i) an isosurface plot along with the predicted trajectory (red dashed line) is depicted.}
  \label{Fig0}
\end{figure*}

\begin{figure*}
\centering
\includegraphics[width=\textwidth]{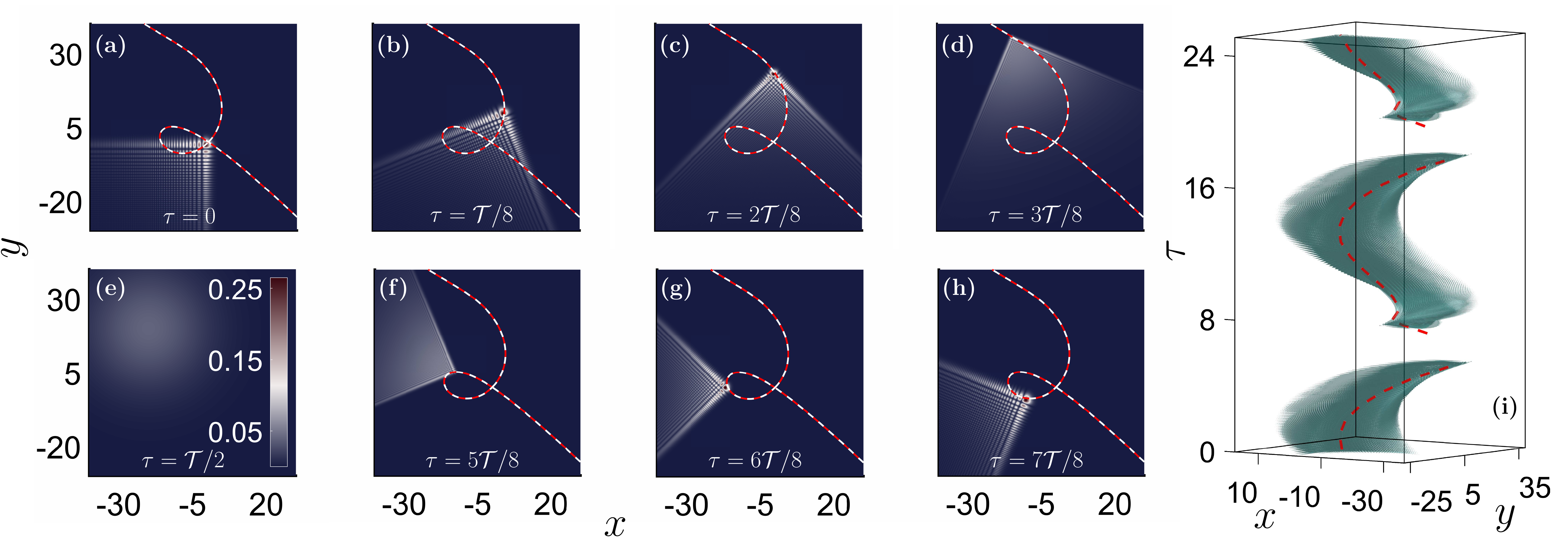} 
\caption{Same as in Fig.~\ref{Fig0} with $\beta_1=\beta_2=5$.}
\label{Fig1}
\end{figure*}

\begin{figure*}[!ht]
\centering
\includegraphics[width=\textwidth]{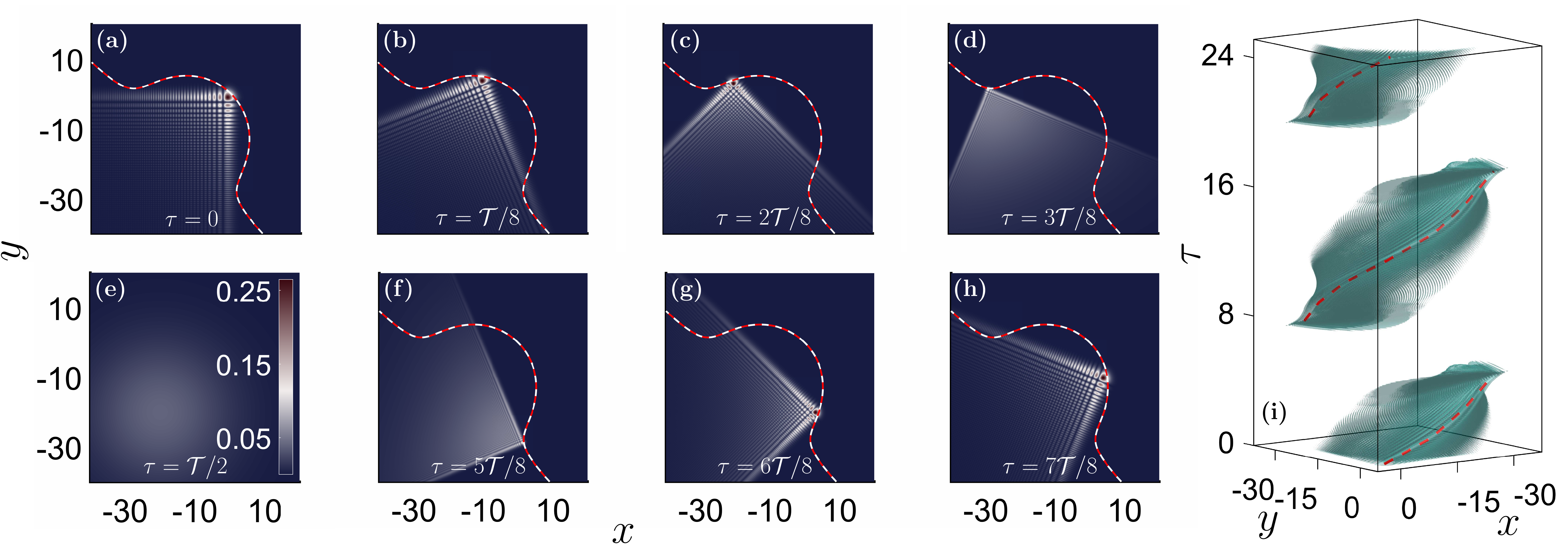}
\caption{Same as in Fig.~\ref{Fig0} with $\beta_1=-5$ and $\beta_2=5$.}
\label{Fig2}
\end{figure*}

\section{Numerical methods}

In the simulations presented in the following sections, we numerically solve Eq.~(\ref{eq:schrodinger}) by utilizing a split-step Fourier method~\cite{yang-siam2010}. We use a second-order Strang-type splitting scheme~\cite{strang-jna1968}, although higher order schemes, such as the Yoshida forth-order splitting~\cite{yoshi-pla1990} can also be applied. The Schr\"odinger type operator on the right hand side of Eq.~(\ref{eq:schrodinger}), can be separated into a diffraction operator $L_D$, an operator that accounts for the parabolic potential $L_P$, and an operator for the orbital angular momentum $L_A=(B/2)L_z$. Note that $L_A$ can be separated into two operators, $L_{A,1}$ and $L_{A,2}$, which are proportional to $x\partial_y$ and $y\partial_x$, respectively. We note that $L_A$ involves the angular coordinate, whereas $L_P$ is a function of the radial coordinate and thus $[L_A,L_P]=0$. By applying a Strang splitting, and using the commutator to reduce the number of operations involved, we obtain
\begin{multline*}
  \psi(\tau+\delta)=e^{(L_D+L_P+L_A)\delta}\psi(\tau)=
  \\
  e^{\frac{L_D\delta}{2}}
  e^{\frac{L_{A,1}\delta}{2}}
  e^{L_{A,2}\delta}
  e^{\frac{L_{A,1}\delta}{2}}
  e^{L_P\delta}
  e^{\frac{L_D\delta}{2}}
  \psi(\tau)
  +\mathcal O(\delta^3)
\end{multline*}
As usual, the diffraction operator is solved by utilizing Fourier transformations
\[
  e^{L_D\delta}\phi =
  \mathcal F^{-1}\{
  e^{-ik^2\delta/2}
  \mathcal F\{\psi
  \}
  \},
\]
while the operator for the parabolic potential is directly integrated in real space resulting to a parabolic phase. The integration of the $L_{A,1}$ operator is obtained by applying a Fourier transformation along the $y$-coordinate, $\mathcal F_y$, leading to
\[
  e^{L_{A,1}\delta}\psi =
  \mathcal F_y^{-1}\{e^{-iBxk_y\delta/2}
  \mathcal F_y\{\psi\}\}.
\]
A similar procedure, involving a Fourier transformation along the $x$-coordinate, is also utilized for the $L_{A,2}$ operator.

\section{Airy beams}

We consider the evolution of an electron wave packet, shaped in the form of a two-dimensional finite energy Airy beam. The initial condition is given by
\begin{equation}
  \psi_0(x,y)=\Ai(\kappa_1 x)\Ai(\kappa_2 y)
  e^{\alpha ( x + y )}
  e^{i(\beta_1 x +\beta_2 y)},
  \label{eq:Airy}
\end{equation}
where $\kappa_1$, $\kappa_2$ are parameters controlling the beam width as well as the transverse acceleration, $\beta_1$, $\beta_2$ are the initial transverse velocities in the $x$- and $y$- direction respectively, and $\alpha$ is an exponential truncation constant that makes the energy of the Airy beam finite. The solution of Eq.~(\ref{eq:schrodinger})  subject to the initial condition given by Eq.~(\ref{eq:Airy}) can be obtained by utilizing the integral representation of Eq.~(\ref{Sol2}). We first note that Eq.~(\ref{Sol2}) can be separated into two one-dimensional integrals. Specifically, we define $f_1(\xi)=\Ai(\kappa_1\xi)e^{\alpha\xi}$, $g_1(\xi)=e^{i v\xi^2}$, $f_2(\eta)=\Ai(\kappa_2\eta)e^{\alpha\eta}$ and $g_2(\eta)=e^{i v\eta^2}$ and thus
\begin{multline}
  \psi= \frac{w}{2\pi i} \int_{\mathbb R}
  f_1(\xi)g_1(\xi)
  e^{-i(S_1-\beta_1)\xi}
  \dif \xi
  \times
  \\
  \int_{\mathbb R}f_2(\eta)g_2(\eta)
  e^{-i(S_2-\beta_2)\eta}\dif\eta,
\label{Sol3}
\end{multline}
We are going to examine the solution of the first integral in Eq.~(\ref{Sol3}), while the second one can be obtained in a similar fashion. The Fourier transform of $f_1(\xi)$ is a Gaussian with a cubic phase
\begin{equation*}
  \mathcal F\{f_1(\xi)\}=
  \frac{1}{\kappa_1}
  e^{\frac{\alpha^3}{3\kappa_1^3}
    -\frac{\alpha k^2}{\kappa_1^3}
    +\frac{i}{3\kappa_1^3}(k^3-3\alpha^2 k)
  },
\end{equation*}
while the Fourier transform of $g_1(\xi)$ is a parabolic phase
\begin{equation}
  \mathcal F\{g_1(x)\}=
  \sqrt{i\frac{\pi}{v}}
  e^{-\frac{i}{4v}k^2}.
\end{equation}
Then, we compute the convolution of the two Fourier transforms with respect to the spatial frequency $k=S_1-\beta_1$ leading to
\begin{multline*}
  \mathcal F\{f_1\} * \mathcal F\{ g_1\}
  =\frac{1}{ 2\pi \kappa_1}\sqrt{i\frac{\pi}{v}}
  e^{\frac{\alpha^3}{3\kappa_1^3}-\frac{i}{4v}(S_1-\beta_1 )^2}
  \\
  \times \int_{-\infty}^{\infty}
  e^{i\frac{u^3}{3\kappa_1^3}-(\frac{\alpha}{\kappa_1^3}+\frac{i}{4v})u^2
    +[-\frac{i\alpha^2}{\kappa_1^3}+ \frac{i}{2v} (S_1-\beta_1 )]u}\dif u.
\end{multline*}
The above integral involves a cubic phase, and thus can be expressed in terms of Airy functions. Taking the product of the solutions we obtain
\begin{equation} 
  \psi=\frac{w}{2v}
  \prod_{j=1}^2
  \Ai(\kappa_j^4 W_j)
  e^{
    \kappa_j^6 (G_j W_j-\frac{G_j^3}{3})
    -\frac{i}{4v}(S_j-\beta_j)^2
    +\frac{\alpha^3}{3\kappa_j^3}
  },
\label{Sols}
\end{equation}
where the arguments $W_j$ in the Airy functions are given by 
\begin{align}
  W_1(x,y,\tau)  &=  \frac{1}{\kappa_1^3}\left(x+\frac{y}
           {\cot \frac{B\tau}{2}}-\frac{\beta_1}{2v} \right)-\frac{1}{16 v^2}+\frac{i\alpha}{2\kappa_1^3 v}, \label{eq:W1}\\ 
  W_2(x,y,\tau)  &= \frac{1}{\kappa_2^3} \left(y-\frac{x}{\cot \frac{B\tau}{2}}-\frac{\beta_2}{2v} \right) - \frac{1}{16 v^2}+ \frac{i\alpha}{2\kappa_2^3 v}, \label{eq:W2}
\end{align}
and 
\begin{equation*}
  G_j= \frac{\alpha}{\kappa_j^3}+\frac{i}{4v}.
\end{equation*}
From Eqs.~(\ref{eq:W1})-(\ref{eq:W2}) we see that, as expected, the solution is not reparable in terms of its arguments $x$ and $y$. 
By setting the real part of the arguments of the Airy functions to zero, we can determine the trajectory $(x_c(\tau),y_c(\tau))$ of the accelerating beam 
\begin{align}
  x_c & =\frac{2B\beta_1 C(\tau) +(\kappa_1^3-2B\beta_2)C^2(\tau)-\kappa_2^3 C^3(\tau)}{B^2 (1+C^2(\tau) )},
        \label{tr1}
  \\
  y_c & =\frac{2B\beta_2 C(\tau) +(\kappa_2^3+2B\beta_1)C^2(\tau)+\kappa_1^3 C^3(\tau)}{B^2 (1 +C^2(\tau) )},
        \label{tr2}
\end{align}
where the time dependence arises from the term $C(\tau)=\tan(B\tau/2)$. The above formulas clearly highlight the periodicity of the trajectory due to the presence of the magnetic field, with period $\mathcal{T}=2\pi/B$. Interestingly, at half the time between successive periods $\tau_{p,l}=(2l+1)\mathcal{T}/2$, both $x(\tau)$ and $y(\tau)$ become infinite. This happens because our calculations for the trajectory are based on the infinite energy Airy beam ($\alpha=0$) and, thus, does not account for the exponential apodization. In finite energy Airy beams the truncation does not allow for the presence of such infinities. Moreover, for $\tau=\tau_{p,l}$ the parameter $v$ is zero, and the exact solution of Eq.~(\ref{Sols}) can be significantly simplified. In particular, Eq.~(\ref{Sol2}) becomes the two-dimensional Fourier transform of the initial condition 
\begin{equation}
  \psi=\frac{Bs}{4i\pi}
  \prod_{j=1}^2
  \frac{1}{\kappa_j}
  e^{
    \frac{\alpha^3}{3\kappa_j^3}
    -\frac{\alpha}{\kappa_j^3} Q_j^2
    +\frac{i}{3\kappa_j^3}Q_j^3
    -i\frac{\alpha^2}{\kappa_j^3}Q_j
  },
\label{GSol}
\end{equation}
where $Q_1=(By/2-\beta_1)$, $Q_2=(-Bx/2-\beta_2)$, and $s=1$ or $s=-1$, depending on whether $l$ is odd or even, respectively. Note that the resulting Gaussian wavefunction is centered at $(-2 \beta_2/B, 2\beta_1/B)$ and, thus, is shifted from the origin when the initial velocities $\beta_j$ are nonzero. Furthermore, the width of the Guassian beam along the $j$th direction $(2\kappa_j^3/(B^2\alpha))^{1/2}$ increases as we increase $\kappa_j$ and as we decrease $B$ and $\alpha$.

To understand the motion of the Airy wave packet we utilize a ray decomposition. In the absence of the magnetic field, the Airy beam of Eq.~(\ref{eq:Airy}) consists of rays that propagate both ``forward'' (towards the upper-right quadrant) and ``backwards'' (towards the lower-left quadrant), with each ray having its own velocity direction and magnitude~\cite{kagan-josaa2012}. The acceleration of the Airy beam is the outcome of the increasing velocity of the forward propagating rays, which contribute to the formation of the beam characteristics. On the other hand, the backward propagating rays do not have any contribution to the formation of the caustic for $z>0$ (or equivalently $t>0$). Virtually, these rays contribute to the caustic when $z<0$ (or $t<0$).

In Fig.~\ref{fig:ray_picture_Airy} we utilize the ray Eqs.~(\ref{stp_1})-(\ref{stp_2}), which are derived in Section~\ref{sec:power-law}, to sketch the ray picture profile when an Airy beam without additional linear velocities ($\beta_1=\beta_2=0$) is utilized as an initial condition [see Eq.~(\ref{eq:Airy})]. The beam (or caustic) trajectory is given by Eqs.~(\ref{tr1})-(\ref{tr2}) or by Eqs.~(\ref{tr3})-(\ref{tr4}) with $\nu=2$. As expected, due to the presence of the magnetic field, the rays become circular, with their radius being proportional to the magnitude of the transverse velocity $R=v_T/B$. Assuming that $B>0$, the direction of the center of the circles is on the left side as we transverse the ray. As a result, at the early stage the, originally straight, caustic bends to the left and forms the left fold of the cusp catastrophe shown in Fig.~\ref{fig:ray_picture_Airy}. Interestingly, the rays that initially move to the lower-left, at the end of their periodic motion, combine to form the lower-right fold of the cusp catastrophe.

\begin{figure*}
  \includegraphics[width=\textwidth]{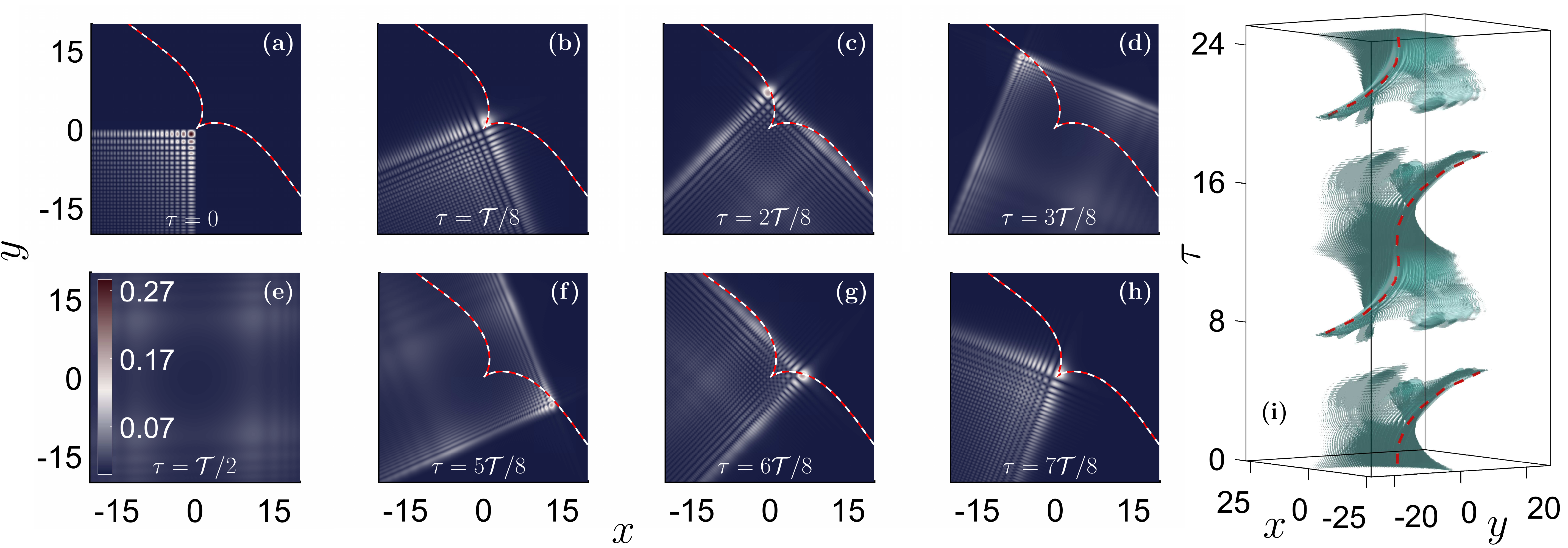}
  \caption{Dynamics of a power-law self-accelerating electron beam with $\delta=1$, $\nu=3/2$, and $\alpha=0.04$ in a constant magnetic field with $B=1/2$. The initial velocities are $\beta_1=\beta_2=0$. Cross sections of the wave amplitude $|\psi|$ are shown on the left panel in multiples of $1/8$th of the period. The colormap shown in (e) is the same for all the cross-sections. In (i) an isosurface plot along with the predicted trajectory (red dashed line) is depicted.}
\label{Fig3}
\end{figure*}

\begin{figure*}
\includegraphics[width=\textwidth]{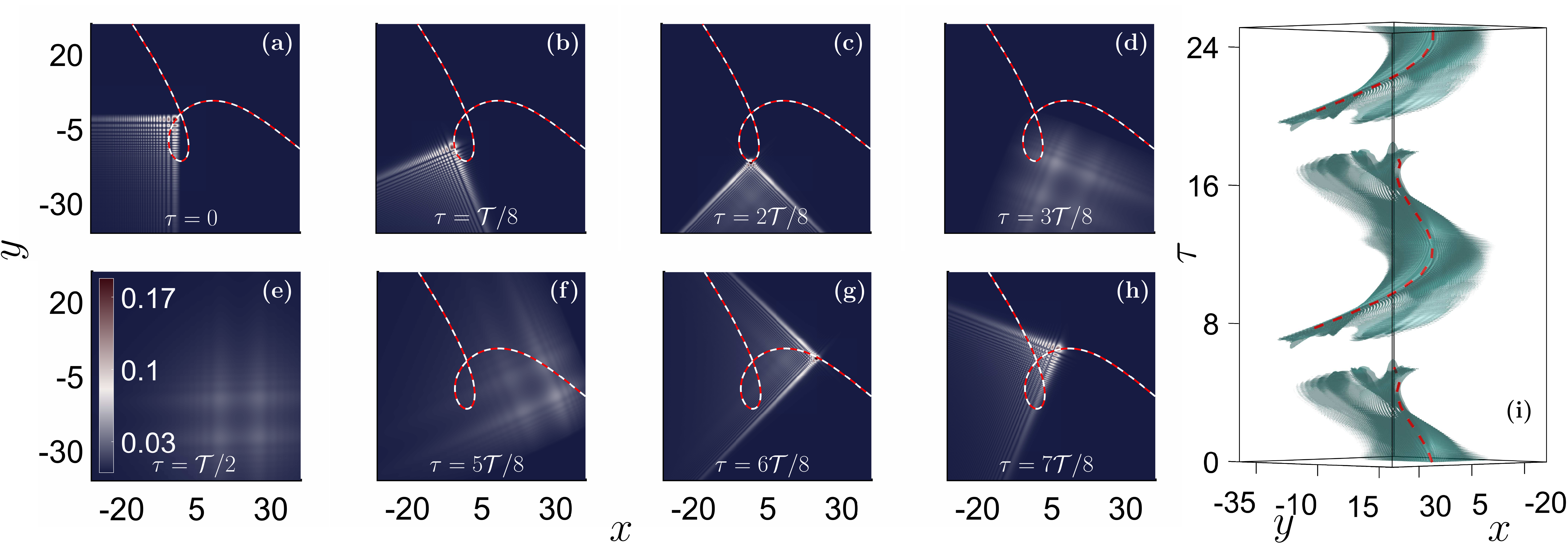} 
\caption{Same as in Fig.~\ref{Fig3} with $\beta_1=\beta_2=-5$.}
\label{Fig4}
\end{figure*}

In Fig.~\ref{Fig0}, we see the dynamics of an exponentially truncated Airy beam given by Eq.~(\ref{eq:Airy}), with an apex
facing towards the upper-right direction of the transverse plane. The initial transverse velocities $\beta_1$ and $\beta_2$ are zero. The main differences, in comparison to the results shown in Fig.~\ref{fig:ray_picture_Airy}, arise from the exponential truncation of the Airy beam. As a result, the beam trajectory in not able to follow the cusp caustic up to infinity. We can see in Fig.~\ref{Fig0}(a)-(d) that the Airy beam initially follows the upper-left fold of the cusp. Even at $\tau=3\mathcal T/8$ the beam trajectory compares well with the theoretical prediction. After that point, diffraction starts to take over and the caustic is effectively destroyed. Exactly at half the period, the wave packet takes a Gaussian profile [see Eq.~(\ref{GSol})]. Due to the wide spectrum of the Airy beam, the maximum amplitude of such a Gaussian Fourier transform is lower as compared to the maximum amplitude along the caustic. After that point, the Airy beam starts to reconstruct along the lower-right fold of the cusp, as we can see in Fig.~\ref{Fig0}(f). Its motion is towards the origin which is reached after each period.

In Figs.~\ref{Fig1}-\ref{Fig2} the Airy beam has an additional linear phase $\beta_1x+\beta_2y$, which results to an initial velocity $(\beta_1,\beta_2)$. The rays due to the isolated motion of this term are going to follow circular trajectories with radius $(\beta_1^2+\beta_2^2)^{1/2}/B$. The combined ray motion due to the Airy beam and the initial velocity is the superposition of the two, as it is also evident from the argument of the Airy beam in Eqs.~(\ref{eq:W1})-(\ref{eq:W2}).
Specifically, in Fig.~\ref{Fig1} the initial velocity points towards the upper-right direction of the transverse plane. When the magnitude of the velocity of the Airy beam is small, then the dominant contribution to the trajectory comes from the linear velocity. This leads to the semicircular part of the trajectory that takes place when $\tau\in[0,\mathcal T/4]\cup[3\mathcal T/4,\mathcal T]$. On the other hand, as $\tau$ approaches $\mathcal T/2$, the velocity of the Airy beam increases and can even become larger than the linear velocity, resulting to the two quasi-linear branches of the trajectory. The parallel shift between these branches arises from the direction of the two velocities, which can be the same (along the upper-left fold) or the opposite (along the lower-right fold). Finally, the loop is the outcome of the change in the direction of the total velocity of the beam. Specifically, at $\tau=5\mathcal T/8$ the wave packet propagates backwards (the Airy velocity is larger than the linear velocity), while at $\tau=3\mathcal T/4$ the wave packet propagates forward (the linear velocity is larger than the Airy velocity).
Finally, in Fig.~\ref{Fig2} the initial linear velocity points towards the upper-left direction and, thus, is perpendicular to the velocity of the Airy beam. Again, the semi-circular part of the trajectory originates from the rays with a dominant velocity due to the linear phase term. Along the diverging upper-left and lower-right parts of the trajectory, the velocity of the Airy beam is stronger than the linear velocity. 


\section{Self-accelerating power-law beams\label{sec:power-law}}

A power-law phase gives rise to an electron beam with a self-accelerating
power-law trajectory~\cite{chrem-ol2011,froeh-oe2011,green-prl2011}. In particular, in the case of one transverse direction, and assuming that $\psi_0=A(x)e^{i\phi(x)}$, the trajectory $x = \delta \tau^\nu$ is generated by the phase factor~\cite{gouts-pra2018}
\begin{equation}
\phi(x)=\frac{- \delta^{1/ \nu} \nu^2 (-x)^{2-1/ \nu}}{(\nu-1)^{1-1/\nu}(2\nu-1)}.
\label{phase}
\end{equation}
Here, we are going to examine the dynamics of such self-accelerating electron beams with power-law trajectories in the presence of a magnetic field. We assume that the initial wave packet can be written as
\begin{equation}
  \psi_0(x,y)=A(x,y)\sin(\phi(x))\sin(\phi(y))e^{i(\beta_1 x+\beta_2 y)},
  \label{eq:icpowerlaw}
\end{equation}
where the phase $\phi$ is given by Eq.~(\ref{phase}). We substitute Eq.~(\ref{eq:icpowerlaw}) to the integral representation of Eq.~(\ref{Sol2}) and utilize a stationary phase method. Using an exponential decomposition of the sine functions and defining by $\Phi$ the phase of the integrand, then from first order phase stationarity ($\Phi_\xi=\Phi_\eta=0$) we derive the ray equations
\begin{align}
  \frac{\dif\phi(\xi)}{\dif\xi}+2v\xi-2v \left(x+\frac{y}{\cot \frac{B\tau}{2}}-\frac{\beta_1}{2v} \right)&=0, \label{stp_1}
  \\
\frac{\dif\phi(\eta)}{\dif\eta}+2v\eta-2v\left(y-\frac{x}{\cot \frac{B\tau}{2}}-\frac{\beta_2}{2v} \right)&=0. \label{stp_2}
\end{align}
In addition, utilizing second order phase stationarity
$\Phi_{\xi\xi}\Phi_{\eta \eta}-\Phi_{\xi \eta}^2=0,$
we find the following implicit relations for the beam trajectory
\begin{equation*}
  x_c=\frac{\delta}{|2v|^\nu}-\frac{y_c}{\cot \frac{B\tau}{2}}+\frac{\beta_1}{2v},\ 
  y_c=\frac{\delta}{|2v|^\nu}+\frac{x_c}{\cot \frac{B\tau}{2}}+\frac{\beta_2}{2v}.
\end{equation*} 
The above formulas can be explicitly solved as a function of time
\begin{align}
  x_c & =\frac{2^\nu B\delta(\chi^2-\chi)+2\beta_1\chi|B\chi|^\nu-2\beta_2|B\chi|^\nu}
     {|B \chi|^{\nu} B(1+\chi^2)},
     \label{tr3}
  \\ 
  y_c & =\frac{2^\nu B\delta(\chi^2+\chi)+2\beta_2\chi|B\chi|^\nu+2\beta_1|B\chi|^\nu}
     {|B\chi|^{\nu} B(1+\chi^2)},
     \label{tr4}
\end{align}
where $\chi=\cot(B\tau/2)$. Note that in the case of a parabolic trajectory ($\nu=2$, Airy beam) from Eqs.~(\ref{tr3})-(\ref{tr4}) with $\delta=\kappa^3/4$ we recover Eqs.~(\ref{tr1})-(\ref{tr2}). 

In our simulations, the amplitude on the input plane is selected as
\[
  A(x,y)=\frac{e^{\alpha(x+y)}}{\pi(xy)^{1/4}},
\]
where $x<0$, and $y<0$.
The results that we obtain, both in terms of ray pictures and wave dynamics, compare qualitatively well to those derived in the case of Airy beams. 
For example, in Fig.~\ref{Fig3} we see the dynamics of a power-law self-accelerating wave with exponent $\nu=3/2$. The wave dynamics are similar with those shown in Fig.~\ref{Fig0}, leading to a cusp-type trajectory. In Fig.~\ref{Fig4}, we have included a linear phase corresponding to a transverse velocity towards the bottom-left, i.e., in a direction opposite to the velocity of the self-accelerating beam. It is interesting to note that, qualitatively, the caustic trajectory looks similar to a mirror image, along the $x=y$ plane, of Fig.~\ref{Fig1}. This happens because the linear velocity field has the same direction as the velocity of the Airy beam along the lower-right fold of the cusp, and the opposite direction along the upper-left fold of the cusp.

\section{CONCLUSIONS}

In conclusion, we have examined the dynamics of electron wave packets of the Airy type, as well as more generic power-law self-accelerating beams, in the presence of an external magnetic field. The qualitative behavior of the such wave packets is unveiled by utilizing a ray picture interpretation. For electron Airy beams, we have analytically derived the dynamics by utilizing the respective Mehler kernel. For generalized beams with power-law phase, we have asymptotically found the ray profile as well as the beam trajectory. Interestingly, the resulting motion of the caustic exhibits a completely different behavior, as compared to the usual circular motion of an electron wave packet.

\section*{ACKNOWLEDGMENTS}
This research is funded by the Greek State Scholarships Foundation
(IKY), project (MIS-5000432).

\hypersetup{urlcolor = blue}
\newcommand{\noopsort[1]}{} \newcommand{\singleletter}[1]{#1}

\end{document}